\documentstyle[aps,tighten,epsf,floats]{revtex}
\def\Jn#1#2#3#4{{#1} {\bf #2}, #3 (#4)}
\def\PRC{{\em Phys. Rev.} C}

\def\PR{{\em Phys. Rev.}}
\def\AP{{\em Ann. Phys. (N.Y.)}}
\def\NC{{\em Nuovo Cimento}}
\def\NCS{{\em Nuovo Cimento Suppl.}}
\def\NP{{\em Nucl. Phys.}}

\def\RPP{{\em Rep. Prog. Phys.}}
\def\IB{{\em ibid.}}

\def\bi{\bibitem}
\def\be{\begin{equation}}
\def\ee{\end{equation}}
\def\bea{\begin{eqnarray}}
\def\eea{\end{eqnarray}}
\newcommand{\A}{{\alpha}}
\newcommand{\B}{{\beta}}

\newcommand{\eqn}[1]{\label{#1}}
\newcommand{\eq}[1]{Eq.~(\ref{#1})}
\newcommand{\eqs}[1]{Eqs.~(\ref{#1})}
\newcommand{\fign}[1]{\label{#1}}
\newcommand{\fig}[1]{Fig.~\ref{#1}}

\newcommand{\NN}{$N\!N$}
\newcommand{\NNN}{$N\!N\!N$}
\newcommand{\NNNN}{$N\!N\!\rightarrow\!N\!N$}
\newcommand{\piN}{$\pi N$}
\newcommand{\piNN}{$\pi N\!N$}

\newcommand{\NNpid}{$N\!N\!\rightarrow\!\pi d$}
\newcommand{\pidpid}{$\pi d\!\rightarrow\!\pi d$}
\newcommand{\pidNN}{$\pi d\!\rightarrow\!N\!N$}

\newcommand{\VOPE}{V_{N\!N}^{\mbox{\protect\scriptsize OPE}}}

\newcommand{\F}{{\cal F}}
\newcommand{\M}{{\cal M}}
\newcommand{\V}{{\cal V}}
\newcommand{\G}{{\cal G}}
\newcommand{\T}{{\cal T}}
\newcommand{\tX}{\tilde{X}}

\newcommand{\bphi}{\bar{\phi}}

\newcommand{\bF}{\bar{F}}
\newcommand{\bB}{\bar{B}}

\newcommand{\GNN}{D_0}
\newcommand{\VNN}{V_{N\!N}}
\newcommand{\TNN}{T_{N\!N}}
\newcommand{\bT}{{\bar{T}}}

\begin{document}

\title{UNIFIED MODEL OF THE RELATIVISTIC \mbox{\boldmath{$\pi N\!N$}}
AND \mbox{\boldmath{$\gamma\pi N\!N$}} SYSTEMS}

\author{B. BLANKLEIDER, A. N. KVINIKHIDZE \footnote{On leave from The
Mathematical Institute of Georgian Academy of Sciences, Tbilisi, Georgia.}}

\address{Department of Physics, The Flinders University of South Australia,\\
Bedford Park, SA 5042, AUSTRALIA\\E-mail: boris.blankleider@flinders.edu.au}

%%%%%%%%%%%%%%%%%%%%%%%%%%%%%%%%%%%%%%%%%%%%%%%%%%%%%%%%%%%%%%
% You may repeat \author \address as often as necessary      %
%%%%%%%%%%%%%%%%%%%%%%%%%%%%%%%%%%%%%%%%%%%%%%%%%%%%%%%%%%%%%%

\maketitle
%\abstract{
\begin{abstract}
We present a unified description of the relativistic \piNN\ and $\gamma \pi
N\!N$ systems where the strong interactions are described non-perturbatively by
four-dimensional integral equations. Our formulation obeys two and three-body
unitarity and is strictly gauge invariant with photons coupled to all possible
places in the strong interaction model. At the same time our formulation is free
from the overcounting problems plaguing four-dimensional descriptions of $\pi
N\!N$ and $\gamma \pi N\!N$ systems.  The accurate nature of our description is
achieved through the use of the recently introduced gauging of equations method.

\end{abstract}
\section{Introduction}

Recently we have introduced a method for incorporating an external
electromagnetic field into any model of hadrons whose strong interactions are
described through the solution of integral equations \cite{gnnn4d,g4d}. The
method involves the gauging of the integral equations themselves, and results in
electromagnetic amplitudes where an external photon is coupled to every part of
every strong interaction diagram in the model. Current conservation is therefore
implemented in the theoretically correct fashion. Initially we applied our
gauging procedure to the covariant three-nucleon problem whose strong
interactions are described by standard four-dimensional three-body integral
equations \cite{gnnn4d,g4d}. More recently we used the same method to gauge the
three-dimensional spectator equation for a system of
three-nucleons \cite{g3d,gnnn3d}.

Here we apply our gauging procedure to the more complicated case of the
covariant \piNN\ system whose four-dimensional integral equations have only
recently been derived \cite{4d,PA4d}. In this way we obtain gauge invariant
expressions for all possible electromagnetic processes of the \piNN\ system,
e.g. pion photoproduction $\gamma d\rightarrow \pi N\! N$, pion
electroproduction $e d\rightarrow e' \pi d$, and because pion absorption is
taken into account explicitly, we also obtain gauge invariant expressions for
processes like deuteron photodisintegration $\gamma d\rightarrow N N$ and
Bremsstrahlung $NN\rightarrow \gamma NN$ that are valid even at energies above
pion production threshold. Included in the electromagnetic processes described
by our model is the especially interesting case of elastic electron-deuteron
scattering. Here the deuteron is described as a bound state of our full \piNN\
model; thus, after gauging, we obtain a rich description of the electromagnetic
form factors of the deuteron with all possible meson exchange currents taken
into account.

\section{Four-Dimensional \mbox{\boldmath{$\pi N\!N$}} Equations}

Since the early 1960's many attempts have been made to formulate few-body
equations for \piNN-like systems using relativistic quantum field theory
\cite{Taylor,Tucciarone,Broido,Avishai4d,AB4d,Haberzettl}. Yet all these
attempts have had theoretical inconsistencies with perturbation diagrams being
either overcounted or undercounted. The first consistent \piNN\ equations were
derived by us only recently \cite{4d} and have the feature of containing
explicit subtraction terms in the kernels of the equations that eliminate all
overcounting. At the same time the undercounting problem is eliminated by the
retention of certain three-body forces. The same \piNN\ equations were later
derived by Phillips and Afnan~\cite{PA4d} using a different method. In this
section we would simply like to restate these equations but in a form that is
particularly convenient for gauging.

\subsection{Distinguishable Nucleon Case}

It is easy to rearrange the four-dimensional \piNN\ equations of Ref.\ 5  into a
convenient form similar to the one used by Afnan and Blankleider~\cite{AB_80} in
a three-dimensional formulation of the \piNN\ system. For the distinguishable
nucleon case we obtain
\be
\T^d = \V^d + \V^d \G_t^d \T^d       \eqn{BS^d}
\ee
where $\T^d$, $\V^d$, and $\G_t^d$ are $4\times 4$ matrices
given by
\be
\T^d=\left(\begin{array}{cc} T^d_{N\!N} & \bT^d_{N} \\
T^d_{N} & T^d \end{array} \right);
\hspace{.3cm}
\V^d=\left( \begin{array}{cc} V^d_{N\!N} & \bar{\F}^d\\
\F^d & G_0^{-1}{\cal{I}} \end{array} \right);
\hspace{.3cm}
\G_t^d=\left(\begin{array}{cc} \GNN & 0 \\
0 & G_0 w^0 G_0 \end{array} \right).  \eqn{AB}
\ee
Here we use the conventions of Ref.\ 13 and refer to the two nucleons as
particles $1$ and $2$, and the pion as particle $3$; also $\lambda=1,2$ labels
the channel where nucleon $\lambda$ and the pion form a two-particle subsystem
with the other nucleon being a spectator, and $\lambda=3$ labels the channel
where the two nucleons form a subsystem with the pion being a spectator.  The
superscript $d$ is used to denote the distinguishable nucleon case on those
symbols that will later be used without the superscript for the case of
indistinguishable nucleons.  \eq{BS^d} is a symbolic equation representing a
Bethe-Salpeter integral equation to be solved for $\T^d$. $\T^d$ consists of
transition amplitudes $T^d_{N\!N}$, $T_{\lambda N}$, $T_{N \mu}$, and
$T_{\lambda\mu}$ ($\lambda$ and $\mu$ are the spectator-subsystem channel
labels) the last three being elements of the matrices $T^d_N$, $\bT^d_{N}$, and
$T^d$, respectively. The physical amplitudes for \NNNN, \NNpid, \pidNN, and
\pidpid\ are then given by
\begin{figure}[t]
%\hspace*{0cm} \epsfxsize=11.5cm\epsffile{v.ps}
\hspace*{2cm} \epsfxsize=13.5cm\epsffile{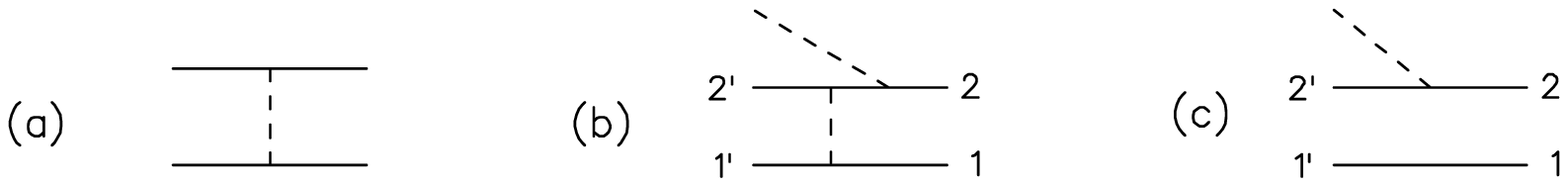}
\vspace{4mm}

\caption{\fign{v} Illustration of quantities making up $V^d_{N\!N}$.  (a)
$\VOPE{^d}$, (b) $B$, and (c) $F_2=f_2d_1^{-1}$.}
\end{figure}
\bea
\begin{array}{ccccccc}
X^d_{N\!N} = T^d_{N\!N}  & ;\hspace{2mm} &
X^d_{dN} = \bar{\psi}_dT_{3N} & ; \hspace{2mm}&
X^d_{Nd} = T_{N3}\psi_d & ; \hspace{2mm}&
X^d_{dd} = \bar{\psi}_d T_{33}\psi_d ,
\end{array}
\eea
respectively, where $\psi_d$ is the deuteron wave function in the presence of a
spectator pion. The kernel $\V^d$ specified in \eq{AB} consists of the following
elements:
\be
V^d_{N\!N} = \VOPE{^d} + \bB^dG_0B^d - \Delta  \eqn{V_NN}
\ee
where $\VOPE{^d}$ is the nucleon-nucleon one pion exchange potential illustrated
in \fig{v}(a), $B^d=B+PBP$ where $B$ is the diagram illustrated in \fig{v}(b)
and $P$ is the nucleon exchange operator, and $\Delta$ is a subtraction term
that eliminates overcounting. $\F^d$ is a $3\times 1$ matrix whose $\lambda$'th
row element is given by
\be
\F^d_\lambda =\sum_{i=1}^2\bar{\delta}_{\lambda i}F_i - B^d . \eqn{F}
\ee
Here $\bar{\delta}_{\lambda i} = 1-\delta_{\lambda i}$ and $F_i=f_id_j^{-1}$
($i,j=1$ or 2 with $i\ne j$) where $f_i$ is the $N_i\rightarrow \pi N_i$ vertex
function and $d_j$ is the Feynman propagator of nucleon $j$. $F_2$ is
illustrated in \fig{v}(c). Note how $B^d$ here plays the role of a subtraction
term. $\bar{\F}^d$ is the $1\times 3$ matrix that is the time reversed version
of $\F^d$ (similarly for other ``barred'' quantities), $G_0$ is the \piNN\
propagator, and ${\cal I}$ is the matrix whose $(\lambda,\mu)$'th element is
$\bar{\delta}_{\lambda,\mu}$.  Finally the propagator term $\G_t^d$ is a
diagonal matrix consisting of the \NN\ propagator $\GNN$, and the $3\times 3$
diagonal matrix $w^0$ whose diagonal elements are $t_1 d_2^{-1}$, $t_2
d_1^{-1}$, and $t^d_3 d_3^{-1}$, with $t_\lambda$ being the two-body t matrix in
channel $\lambda$ (for $\lambda=1$ or $2$, $t_\lambda$ is defined to be the
\piN\ t matrix with the nucleon pole term removed).
\begin{figure}[b]
%\hspace*{1mm} \epsfxsize=11.5cm\epsffile{x.ps}
\hspace*{2cm} \epsfxsize=13.5cm\epsffile{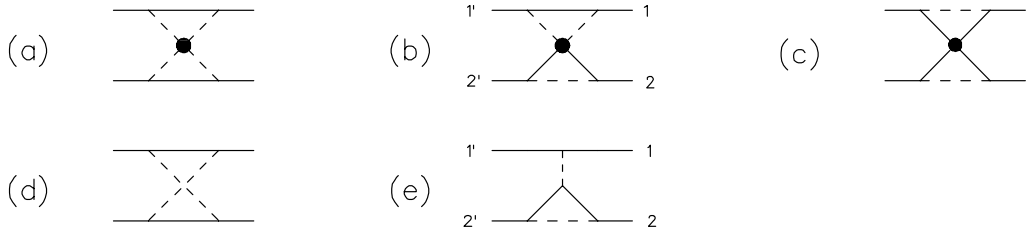}
\vspace{3mm}

\caption{\fign{x} Terms making up the subtraction term $\Delta$.  (a)
$W_{\pi\pi}$, (b) $W_{\pi N}$, (c) $W_{N\!N}$, (d) $X$, and (e) $Y$. The dark
circles represent the following two-body amplitudes: (a) full $\pi\pi$ t-matrix,
(b) one-nucleon irreducible $\pi N$ t-matrix, and (c) full \NN\ t-matrix minus
the \NN\ one-pion-exchange potential.}
\end{figure}

The subtraction term $\Delta$ is defined with the help of \fig{x} as
\be
\Delta = W_{\pi\pi} + W^d_{\pi N} + W_{N\!N} + X + Y^d \nonumber \\
\ee
where $W^d_{\pi N}=W_{\pi N}+PW_{\pi N}P$ and $Y^d=Y+PYP$.

\subsection{Indistinguishable Nucleon Case}

In approaches based on second quantisation in quantum mechanics it is usual to
obtain the scattering equations for identical particles by explicitly
symmetrising the equations of the distinguishable particle case. This procedure
is not justified in the framework of relativistic quantum field theory.
Nevertheless, as we have already derived the \piNN\ equations taking into
account identical particle symmetry right from the beginning \cite{4d}, we can
formally deduce how the above distinguishable nucleon \piNN\ equations need to
be modified in order to get the indistinguishable nucleon case.  With this in
mind, we introduce identical nucleon transition amplitudes defined in terms of
distinguishable nucleon transition amplitudes as
\be
\begin{array}{lll}
\TNN=T^d_{N\!N}A \hspace{1.8cm} & T_{\Delta N}=T_{1N}A \hspace{.8cm} &
T_{dN}=T_{3N}A\\
T_{N\Delta }=\bar{T}_{N1}-\bar{T}_{N2}P
& T_{Nd}=\bar{T}_{N3}A
&T_{\Delta\Delta}=T_{11}-T_{12}P \\
T_{d\Delta }=T_{31}-T_{32}P &
T_{\Delta d}=T_{13}A &T_{dd}=T_{33}A
\end{array} \hspace{.2cm}   \eqn{TA}
\ee
where $A=1-P$ is the antisymmetrising operator. As $\bar{T}_{N1}=P\bar{T}_{N2}P$
and $\bar{T}_{N3}=P\bar{T}_{N3}P$, we can alternatively write $T_{N\Delta
}=A\bar{T}_{N1}$, $T_{Nd}=A\bar{T}_{N3}$, and $T_{d\Delta}=A\bar{T}_{31}$.  Thus
in the transition to the indistinguishable particle case, the original 16
transition amplitudes for distinguishable particles have been reduced to 9
antisymmetrised transition amplitudes. By taking residues of the \piNN\ Green
function for identical nucleons at two-body subsystem poles, one obtains the
following expressions for the physical amplitudes:
\bea
\begin{array}{ccccccc}
X_{N\!N} = \TNN  & ;\hspace{2mm} &
X_{dN} = \bar{\psi}_dT_{dN} & ; \hspace{2mm}&
X_{Nd} = T_{Nd}\psi_d & ; \hspace{2mm}&
X_{dd} = \bar{\psi}_d T_{dd}\psi_d .
\end{array}   \eqn{Xij}
\eea
Using \eqs{TA} in \eq{BS^d} it is easy to show that one again obtains a
Bethe-Salpeter equation
\be
\T = \V + \V\G_t\T       \eqn{BS}
\ee
but where now $\T$, $\V$, and $\G_t$ are
$3\times 3$ matrices given by
\bea
\T &=& \left( \begin{array}{ccc}
\TNN& T_{N\Delta}&T_{Nd} \\
T_{\Delta N}& T_{\Delta\Delta} & T_{\Delta d}\\ 
T_{dN}& T_{d\Delta}&T_{dd}   \end{array}  \right); \hspace{1cm}
\V = \left( \begin{array}{ccc}
V^d_{N\!N}A& A\bar{\F}^d_1 &\bar{\F}^d_3A \\
\F^d_1A & -G_0^{-1}P\hspace{2mm}  & G_0^{-1}A \\ 
\F^d_3A& G_0^{-1}A & 0   \end{array}  \right); \eqn{V} \\
\G_t &=& \left( \begin{array}{ccc}
\frac{1}{2}\GNN& 0 & 0 \\
0&G_0 t_1d_2^{-1}G_0& 0\\ 
0 & 0 & \frac{1}{4}G_0t_3d_3^{-1}G_0   \end{array}  \right) \eqn{G}
\eea
where $t_3=t^d_3A$ is the t matrix for two identical nucleons in the presence
of a spectator pion.

From \eq{V_NN} it follows that
\be
V^d_{N\!N}A=\VOPE - W_{\pi\pi}^R - W_{\pi N}^{LR}- W_{N\!N}^R-X^R - Y^{LR}
+\frac{1}{2}\bB^{LR}G_0 B^{LR}  \eqn{V_NN_A}
\ee
where $\VOPE=\VOPE{^d}A$ and where we have used a superscript $L$ to
indicate that $A$ is acting on the left, and a superscript $R$ to indicate that
$A$ is acting on the right. It is easy to see that $W_{\pi\pi}^R=W_{\pi\pi}^L$, 
$W_{N\!N}^R=W_{N\!N}^L$ and $X^R=X^L$.

We note that the \piNN\ equations for identical nucleons derived in
Ref.\ 6, unlike \eqs{BS}, (\ref{V}) and (\ref{G}), contain no
antisymmetrisation operators $A$ and are therefore not equivalent to our
\piNN\ equations.

\section{\mbox{\boldmath{$\pi N\!N$}} Electromagnetic Transition Currents}

\subsection{Gauging the \piNN\ Equations}

In this section we shall derive expressions for the various electromagnetic
transition currents of the \piNN\ system. To do this we utilise the recently
introduced gauging of equations method \cite{gnnn4d,g4d}. As the gauging
procedure is identical for the distinguishable and indistinguishable particle
cases, we restrict our attention to the \piNN\ system where the nucleons are
treated as indistinguishable.

Direct gauging of \eq{BS} gives
\be
\T^\mu = \V^\mu + \V^\mu \G_t\T + \V\G_t^\mu\T + \V\G_t\T^\mu
\ee
which can easily be solved for $\T^\mu$ giving
\be
\T^\mu = (1+\T\G_t)\V^\mu(1+\G_t\T) + \T \G_t^\mu \T \eqn{T^mu}.
\ee
$\T^\mu$ is a matrix of gauged transition amplitudes $T^\mu_{N\!N}$,
$T^\mu_{N\Delta}$, $T^\mu_{Nd}$, etc. To obtain the physical electromagnetic
transition currents of the \piNN\ system where photons are attached everywhere
it is not sufficient to just gauge the physical \piNN\ amplitudes of \eq{Xij}.
Although this would indeed attach photons everywhere inside the strong
interaction diagrams, it would miss those contributions to the physical
electromagnetic transition currents where the photons are attached to the
external (initial and final state) pions and nucleons.  In order to also include
these external leg contributions it is useful to attach the corresponding
propagators to the $X$-amplitudes of \eq{Xij}:
\bea
\tX_{N\!N} &=&  \GNN X_{N\!N} \GNN \hspace{1cm}
\tX_{dN} =  d_\pi X_{dN} \GNN    \eqn{tX1} \\
\tX_{Nd} &=& \GNN X_{Nd} \, d_\pi  \hspace{1.2cm}
\tX_{dd} = d_\pi X_{dd}\, d_\pi  .  \eqn{tX2}
\eea
The physical electromagnetic transition currents are then obtained by gauging
\eqs{tX1} and (\ref{tX2}) and ``chopping off'' external legs:
\bea
j^\mu_{N\!N} &=& \GNN^{-1}\tX_{N\!N}^\mu \GNN^{-1}\hspace{1cm}
j^\mu_{dN} = d_\pi^{-1} \tX_{dN}^\mu \GNN^{-1} \eqn{X^mu1}\\
j^\mu_{Nd} &=& \GNN^{-1}\tX_{Nd}^\mu\, d_\pi^{-1}\hspace{1.3cm}
j^\mu_{dd} = d_\pi^{-1} \tX_{dd}^\mu \,d_\pi^{-1}.  \eqn{X^mu2}
\eea
Using \eqs{Xij} we obtain that
\bea
j^\mu_{N\!N} &=& \GNN^{-1}D_0^\mu\TNN
+ \TNN D_0^\mu\GNN^{-1} + \TNN^\mu \eqn{j_NN^mu}\\
j^\mu_{dN} &=& \bphi_d^\mu \GNN T_{dN} + d_\pi^{-1}\bphi_d G_0^\mu T_{dN}
+ \bphi_d \GNN T^\mu_{dN} \nonumber \\
&+& \bphi_d \GNN T_{dN} \GNN^\mu \GNN^{-1}\eqn{j_dN^mu}\\
j^\mu_{Nd} &=& T_{Nd}\GNN \phi_d^\mu  + T_{Nd}G_0^\mu \phi_d d_\pi^{-1}
+  T^\mu_{Nd}  \GNN\phi_d \nonumber \\
&+& \GNN^{-1} \GNN^\mu T_{Nd} \GNN \phi_d  \eqn{j_Nd^mu} \\
j^\mu_{dd} &=& \bphi_d^\mu \GNN T_{dd}\GNN\phi_d
+ d_\pi^{-1}\bphi_d \GNN^\mu T_{dd}\GNN\phi_d \nonumber \\
&+& \bphi_d \GNN T_{dd}^\mu \GNN\phi_d
+ \bphi_d \GNN T_{dd}\GNN^\mu\phi_d d_\pi^{-1}\nonumber \\
&+& \bphi_d \GNN T_{dd}\GNN\phi_d^\mu \eqn{j_dd^mu}
\eea
where $\phi_d$ is the deuteron bound state vertex function defined by the
relation $\psi_d=d_1d_2\phi_d$. Note that $\phi_d^\mu$ consists of
contributions where the photon is attached everywhere inside the deuteron bound
state, and is determined by gauging the two-nucleon bound state equation for
$\phi_d$ \cite{gnnn4d,g4d}.

That the above equations are gauge invariant is evident from the fact that we
have formally attached the photon to all possible places in the strong
interaction model. The gauge invariance of our equations also follows from a
strict mathematical proof; however, as this proof is essentially identical to
the one given for the \NNN\ system \cite{g4d}, we shall not repeat it here.

\subsection{Alternative Form of the \piNN\ Equations}

Although the preceding discussion solves the problem of gauging the \piNN\
system in a straightforward way, the expression obtained to calculate the gauged
transition amplitudes, \eq{T^mu}, is not the most convenient for numerical
calculations. The disadvantage of \eq{T^mu} is that it utilises a Green function
$\G_t$ which contains two-body t matrices. The presence of these t matrices
makes the evaluation of $\T^\mu$ unnecessarily complicated.  We therefore
present an alternative form of the \piNN\ equations which uses a ``free'' Green
function which contains no two-body interactions and which leads to simpler
expressions for the \piNN\ electromagnetic transition currents.

The \piNN\ equations \eqs{BS}, (\ref{V}) and (\ref{G}), can be written in the
form
\be \left(
\begin{array}{cc} \TNN & \bT_{N} \\
T_{N} & T \end{array} \right)= \left( \begin{array}{cc}
\VNN & \bar{\F} \\
\F & L G_0^{-1} \end{array} \right)\left[ 1+ \left(
\begin{array}{cc} \frac{1}{2}\GNN & 0 \\
0 & G_0tG_0 \end{array} \right) \left(
\begin{array}{cc} \TNN & \bT_{N} \\
T_{N} & T \end{array} \right) \right].  \eqn{KB}
\ee
where $\VNN=V^d_{N\!N}A$ is given by \eq{V_NN_A}, and
\bea
\F &=&\left( \begin{array}{c}
\F^d_1A \\ \F^d_3A \end{array}  \right)
=\left[ \begin{array}{c} (F_2-B^d)A \\
(F_1+F_2-B^d)A \end{array}  \right], \\
\bar{\F} &=&\left(A\bar{\F}^d_1\hspace{2mm} A\bar{\F}^d_3\right)
= \left[A(\bar{F}_2-\bB^d)\hspace{3mm} 
A(\bF_1+\bF_2-\bB^d)\right],\\
L&=&\left( \begin{array}{cc}
-P  & A \\ 
A & 0   \end{array}  \right), \hspace{.7cm}t=\left( \begin{array}{cc}
t_1d_2^{-1}  &  0 \\ 
0    & \frac{1}{4}t_3d_3^{-1}  \end{array}  \right).  \eqn{gkb3}
\eea
With the view of gauging Green function versions of \piNN\ transition
amplitudes, we introduce the Green function matrix appropriate to \eq{KB}:
\bea
\G &=& \left(\begin{array}{cc} G_{N\!N} & \bar{G}_{N} \\
G_{N} & G \end{array} \right)
\equiv
\left( \begin{array}{ccc}
G_{N\!N}& G_{N\Delta}&G_{Nd} \\
G_{\Delta N}& G_{\Delta\Delta} & G_{\Delta d}\\ 
G_{dN}& G_{d\Delta}&G_{dd}  \end{array}  \right) \nonumber \\
&=&\left(\begin{array}{cc} A\GNN & 0 \\
0 & 0 \end{array} \right)+\left(
\begin{array}{cc} \GNN & 0 \\
0 & G_0 \end{array} \right)
\left(\begin{array}{cc} \TNN & \bT_{N} \\ 
T_{N} & T \end{array} \right)
\left(\begin{array}{cc} \GNN & 0 \\
0 & G_0 \end{array} \right) . \eqn{G1}
\eea
The inhomogeneous term is chosen so that $G_{N\!N}$ corresponds
exactly to the full Green function for \NN\ scattering.  Then it can be shown
that $\G$ satisfies the equation
\be
\G=\left( \begin{array}{cc}
A\GNN & 0 \\
0  & L G_0 \end{array} \right)
\left[1+\left( \begin{array}{cc}
\frac{1}{4}\VNN-
\frac{1}{4}\bar{\Lambda}L G_0\Lambda& 
\hspace{2mm}\frac{1}{2}\bar{\Lambda} \\
\frac{1}{2}\Lambda & \hspace{2mm}t \end{array} \right)\G\right]
\eqn{G2}
\ee
where $\Lambda$ and $\bar{\Lambda}$ are given by
\be
\Lambda = \left[\begin{array}{c}
\left(F_1-{\textstyle\frac{1}{2}}B^d\right)A \\
{\textstyle -\frac{1}{2}}B^d
\end{array} \right] ; \hspace{.5cm}
\bar{\Lambda} =
\left[A\left(\bF_1-{\textstyle\frac{1}{2}}\bB^d\right)
\hspace{3mm}
{\textstyle -\frac{1}{2}}\bB^d\right]. \eqn{Lam}
\ee
The essential feature of \eq{G2} is that it is written in terms of an effective
``free'' Green function matrix
\be
\G_0 = \left( \begin{array}{cc}
A\GNN & 0 \\
0  & L G_0 \end{array} \right)
\ee
which does not involve two-body interactions. For this reason \eq{G2} is ideal
for the purposes of gauging.

\subsection{Gauging the Alternative Form of the \piNN\ Equations}

In terms of the elements of $\G$, the Green function versions of the
physical amplitudes (defined in \eq{tX2}) are given by
\bea
\begin{array}{ccccccc}
\tX_{N\!N} = G_{N\!N}  & ;\hspace{2mm} &
\tX_{dN} = \bar{\phi}_dG_{dN} & ; \hspace{2mm}&
\tX_{Nd} = G_{Nd}\phi_d & ; \hspace{2mm}&
\tX_{dd} = \bar{\phi}_d G_{dd}\phi_d .
\end{array}
\eea
After gauging, these equations give
\bea
\tX_{N\!N}^\mu &=& G_{N\!N}^\mu \hspace{2.5cm}
\tX_{dN}^\mu = \bar{\phi}_d^\mu G_{dN}+\bar{\phi}_dG_{dN}^\mu \eqn{tX^mu1}\\
\tX_{Nd}^\mu &=& G_{Nd}^\mu\phi_d+G_{Nd}\phi_d^\mu\hspace{.7cm}
\tX_{dd}^\mu = \bar{\phi}_d^\mu G_{dd}\phi_d+\bar{\phi}_d G_{dd}^\mu\phi_d
+\bar{\phi}_d G_{dd}\phi_d^\mu.\hspace{.8cm} \eqn{tX^mu4}
\eea
The \piNN\ electromagnetic transition currents $j^\mu_{\A\B}$ are then
determined by \eqs{X^mu1} and \eqs{X^mu2}.

To determine the quantities $G_{\A\B}^\mu$ in \eqs{tX^mu1} and (\ref{tX^mu4})
we need to derive the expression for $\G^\mu$ by gauging \eq{G2}. Defining
\be
\V_t=\left( \begin{array}{cc}
\frac{1}{4}\VNN-
\frac{1}{4}\bar{\Lambda}L G_0\Lambda& 
\hspace{2mm}\frac{1}{2}\bar{\Lambda} \\
\frac{1}{2}\Lambda & \hspace{2mm}t \end{array} \right)
\ee
\eq{G2} can be written as
\be
\G = \G_0+\G_0 \V_t \G.   \eqn{G3}
\ee
Gauging this equation and solving for $\G^\mu$ gives
\be
\G^\mu=(1+\G\V_t)\G_0^\mu(1+\V_t\G)+\G\V_t^\mu\G.   \eqn{G^mu_tmp}
\ee
To simplify this equation we cannot use \eq{G3} to write
$1+\V_t\G=\G_0^{-1}\G$ and $1+\G\V_t=\G\G_0^{-1}$ since $L$ is
singular so that the inverse $\G_0^{-1}$ does not exist. Instead we use the
fact that
$L=L\Omega L$, where
\be
\Omega= \left( \begin{array}{cc}
-\frac{1+P}{2}  & \frac{1}{2} \\ 
\frac{1}{2} & -\frac{1}{4}   \end{array}  \right)
\ee
which allows us to write
\be
\G_0^\mu = \G_0 \M^\mu \G_0  \eqn{G_0^mu}
\ee
where
\be
\M^\mu=\left( \begin{array}{cc}
\frac{1}{2}A\GNN^{-1}\GNN^\mu \GNN^{-1} & 0 \\
0  & \Omega G_0^{-1}G_0^\mu G_0^{-1} \end{array} \right).
\ee
Using \eq{G_0^mu} in \eq{G^mu_tmp} gives us a compact expression for $\G^\mu$:
\be
\G^\mu = \G\left( \M^\mu + \V_t^\mu \right) \G.    \eqn{G^mu}
\ee
It is easy to see that the term $\M^\mu$ corresponds to photon
coupling in the impulse approximation while
\begin{figure}[t]
\hspace{.5cm} \epsfxsize=16.5cm\epsffile{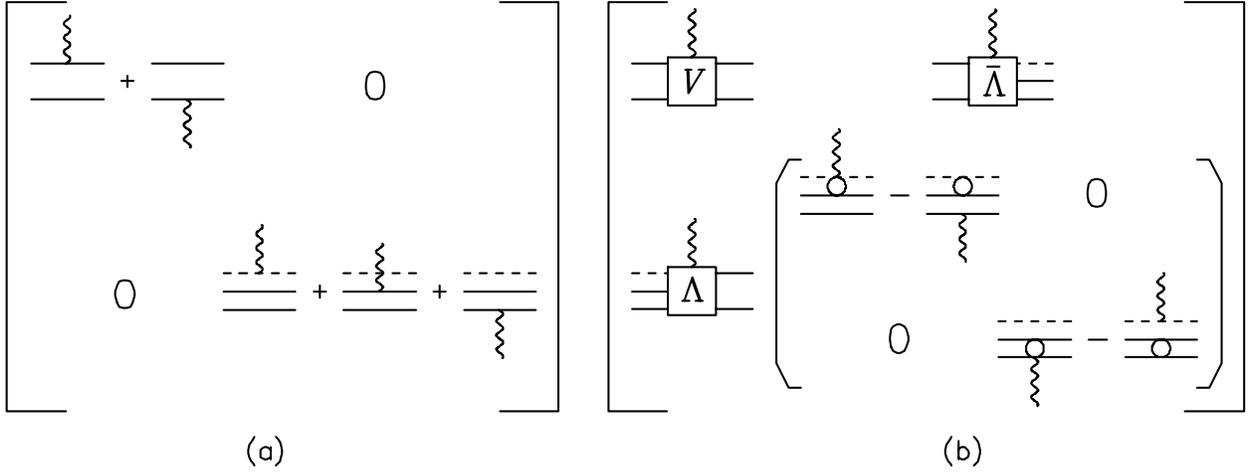}
\vspace{3mm}

\caption{\fign{gmu} Graphical representation of (a) $\M^\mu$ and
(b) $\V_t^\mu$ which enter into the expression for the gauged Green function
matrix $\G^\mu$ of \protect\eq{G^mu}. Constant factors and effects of operator
$A$ and matrix $\Omega$ have been ignored in this illustration.}
\end{figure}
$\V_t^\mu$ corresponds to the interaction currents and consists of
the elements $V^\mu=(\VNN-\bar{\Lambda}LG_0\Lambda)^\mu$, $\bar{\Lambda}^\mu$,
$\Lambda^\mu$ and $t^\mu$. It is important to note that the diagonal elements of
matrix $t^\mu$ are both of the form
\be
\left(t_id_j^{-1}\right)^\mu = t_i^\mu d_j^{-1}+ t_i \left(d_j^{-1}\right)^\mu
=t_i^\mu d_j^{-1} - t_i\Gamma_j^\mu
\ee
where $\Gamma_i^\mu = d_i^{-1} d_i^\mu d_i^{-1}$ is the electromagnetic vertex
function of the nucleon and the last equality follows from the fact that
$\left(d_j^{-1}d_j\right)^\mu=0$. Thus the diagonal elements of $t^\mu$ involve
new subtraction terms $t_i\Gamma_j^\mu$ whose origin does not lie in the
subtraction terms of the strong interaction \piNN\ equations, but rather in the
gauging procedure itself. Analogous subtraction terms arise in the three-nucleon
problem whose strong interaction equations have no subtraction
terms \cite{gnnn4d,g4d}. Similar subtraction terms will arise in the gauging of
$F_1$ and $\bF_1$. The graphical representation of $\M^\mu$ and $\V_t^\mu$ is
given in \fig{gmu}.

\eq{G^mu} can be used to determine all the possible electromagnetic transition
currents of the \piNN\ system. An especially interesting use of \eq{G^mu} is to
study the electromagnetic properties of bound states of the \piNN\ system. It is
certainly expected that the strong interaction \piNN\ model under discussion
admits a bound state corresponding to the physical deuteron. In this case a
solution will exist to the homogeneous version of \eq{BS}:
\be
\Phi = \V\G_t\Phi   \eqn{Phi}
\ee
where $\Phi = \left(\begin{array}{ccc} \Phi_N & \Phi_\Delta &
 \Phi_d \end{array}\right)^T$.
Here $\Phi_N$ is the usual deuteron vertex function describing the $d\rightarrow
N N$ transition, while $\Phi_\Delta$ and $\Phi_d$ are somewhat unusual in that
they describe transitions to clustered \piNN\ states: $d\rightarrow (\pi N) N$
and $d\rightarrow (N N) \pi$ respectively. Comparing
\eq{G1} and \eq{G2} it is seen that $\G$ has a pole at the deuteron mass
$M_d$:
\be
\G \sim i \frac{\Psi\bar{\Psi}}{P^2-M_d^2}\hspace{1cm}\mbox{as}\hspace{.5cm}
P^2\rightarrow M_d^2    \eqn{G_pole} 
\ee
where $P$ is the total four-momentum of the system,
and where $\Psi$ satisfies the equation
\be
\Psi=\G_0\V_t\Psi.   \eqn{Psi}
\ee
Clearly $\Psi$ is related to $\Phi$ by the equation
$
\Psi = \left(\begin{array}{cc} \GNN & 0 \\
0 & G_0 \end{array}\right) \Phi
$
so that either of the equations (\ref{Phi}) or (\ref{Psi}) can be used to
determine $\Psi$.
Taking the left and right residues at the deuteron bound state poles of
\eq{G^mu} we obtain the bound state electromagnetic current
\be
j^\mu = \bar{\Psi}\left(\M^\mu+\V_t^\mu\right)\Psi     \eqn{j^mu}
\ee
which describes the electromagnetic properties of the deuteron whose internal
structure is described by the present \piNN\ model. \eq{j^mu} provides a very
rich description of the internal electromagnetic structure of the deuteron with
all possible meson exchange currents being taken into account in a gauge
invariant way. In view of the accuracy of this model which is based on meson and
baryon degrees of freedom, a comparison of the deuteron electromagnetic form
factors (easily extracted from $j^\mu$) with experiment
should prove to be very interesting.
\section*{Acknowledgements}
The authors would like to thank the Australian Research Council for their
financial support.

\end{document}